\documentclass[a4paper]{article}

\usepackage{INTERSPEECH2022}
\usepackage{graphicx}
\usepackage{subfigure}
\usepackage{latexsym}
\usepackage{url}
\usepackage{cite}

\title{Speaking-Rate-Controllable HiFi-GAN Using Feature Interpolation}
\name{Detai Xin$^1$, Shinnosuke Takamichi$^1$, Takuma Okamoto$^2$, Hisashi Kawai$^2$, Hiroshi Saruwatari$^1$}
\address{
  $^1$Graduate School of Information Science and Technology, The University of Tokyo, Japan.\\
  $^2$National Institute of Information and Communications Technology, Japan.}
\email{\{detai\_xin, shinnosuke\_takamichi\}@ipc.i.u-tokyo.ac.jp}

\begin{document}
\setlength\abovecaptionskip{6pt} % 図とキャプションの間のマージン

\maketitle
\begin{abstract}
This paper presents a speaking-rate-controllable HiFi-GAN neural vocoder.
Original HiFi-GAN is a high-fidelity, computationally efficient, and tiny-footprint neural vocoder.
We attempt to incorporate a speaking rate control function into HiFi-GAN for improving the accessibility of synthetic speech.
The proposed method inserts a differentiable interpolation layer into the HiFi-GAN architecture.
A signal resampling method and an image scaling method are implemented in the proposed method to warp the mel-spectrograms or hidden features of the neural vocoder.
We also design and open-source a Japanese speech corpus containing three kinds of speaking rates to evaluate the proposed speaking rate control method.
Experimental results of comprehensive objective and subjective evaluations demonstrate that 1) the proposed method outperforms a baseline time-scale modification algorithm in speech naturalness, 2) warping mel-spectrograms by image scaling obtained the best performance among all proposed methods, and 3) the proposed speaking rate control method can be incorporated into HiFi-GAN without losing computational efficiency.
\end{abstract}

%1
\section{Introduction}
Speaking rate control is a speech processing task of changing the speaking rate of a given speech~\cite{src_4,src_0,src_1,src_2,mpm}.
This technology has been widely applied in many real-world applications.
Since speaking rate changes speech intelligibility and accessibility, users may need different speaking rates depending on the situation.
For instance, a high speaking rate may be needed by visually-impaired people~\cite{blind,bragg2018large} and students who take online courses~\cite{dodson2021will} since they are used to listening to speech with a higher speed.
Instead, people who are in an environment with high reverberation~\cite{nabvelek1989reverberant} or people who have aphasia~\cite{hux2020effect} and elderly adults~\cite{sr,rudner2011working} may need a low speaking rate to understand the content of speech.

The most prevailing method for speaking rate control is time-scale modification (TSM) algorithms \cite{driedger2016review, muller2015fundamentals} based on waveform warping.\footnote{Although speaking rate control can also be realized by source filter vocoders~\cite{straight,world} and time-domain pitch synchronous
overlap-add (TDPSOLA)~\cite{tdpsola,sinu,ahm}, the
synthesis quality of these methods is highly dependent on the
accuracy of fundamental frequency analysis~\cite{ndf,harvest} and pitch marking~\cite{pitch_marking}.}
Basically, such an algorithm first decomposes a waveform into several frames, then warps each frame to the target rate and finally reconstructs the waveform.
However, although TSM algorithms have acceptable quality and efficiency, their performance is still not satisfactory.
Moreover, recent advances in applying deep neural networks (DNNs) for speech synthesis like neural vocoder~\cite{wn_vocoder,wavernn,waveglow,valin2019lpcnet,kumar2019melgan,pwg,kong2020hifi} demonstrate superior performance in synthesizing speech with high fidelity and efficiency, but since TSM algorithms modify time-domain waveforms directly, it is difficult to combine a DNN-based speech synthesis model with a TSM algorithm.
Therefore, a powerful and efficient speaking rate control method that can be seamlessly implemented in DNN-based speech synthesis models becomes necessary.
A DNN-based speaking rate control method with multi-speaker WaveNet vocoder~\cite{si_wn_vocoder} had been initially provided and it outperformed the conventional TSM-based method and source-filter vocoder~\cite{okamoto2022spcom}.
However, the inference speed of the method was quite slow due to the auto-regressive structure and the huge size of the WaveNet model~\cite{wn_vocoder}.
Most recently, Morrison {\it et al.} \cite{morrison2021neural} proposed a speaking rate control method using LPCNet for real-time inference~\cite{valin2019lpcnet}.
However, their method was mainly designed for pitch shifting, and the neural vocoder used in their work is different from the one used in this work.

%Another issue in evaluation of speaking rate control is that we do not know the ground-truth for speed-controlled speech.
Another issue in the evaluations of previous work is a lack of ground-truth speaking-rate-controlled speech.
The goal of the control is to synthesize speech as if the speaker has uttered the speech with the specified speaking rate.
However, since past studies using existing corpora~\cite{okamoto2022spcom, morrison2021neural} always compared speaking-rate-controlled speech with original speech, we cannot state how far those control methods are from the goal.

\begin{figure}[t]
  \centering
  \includegraphics[width=0.98\linewidth]{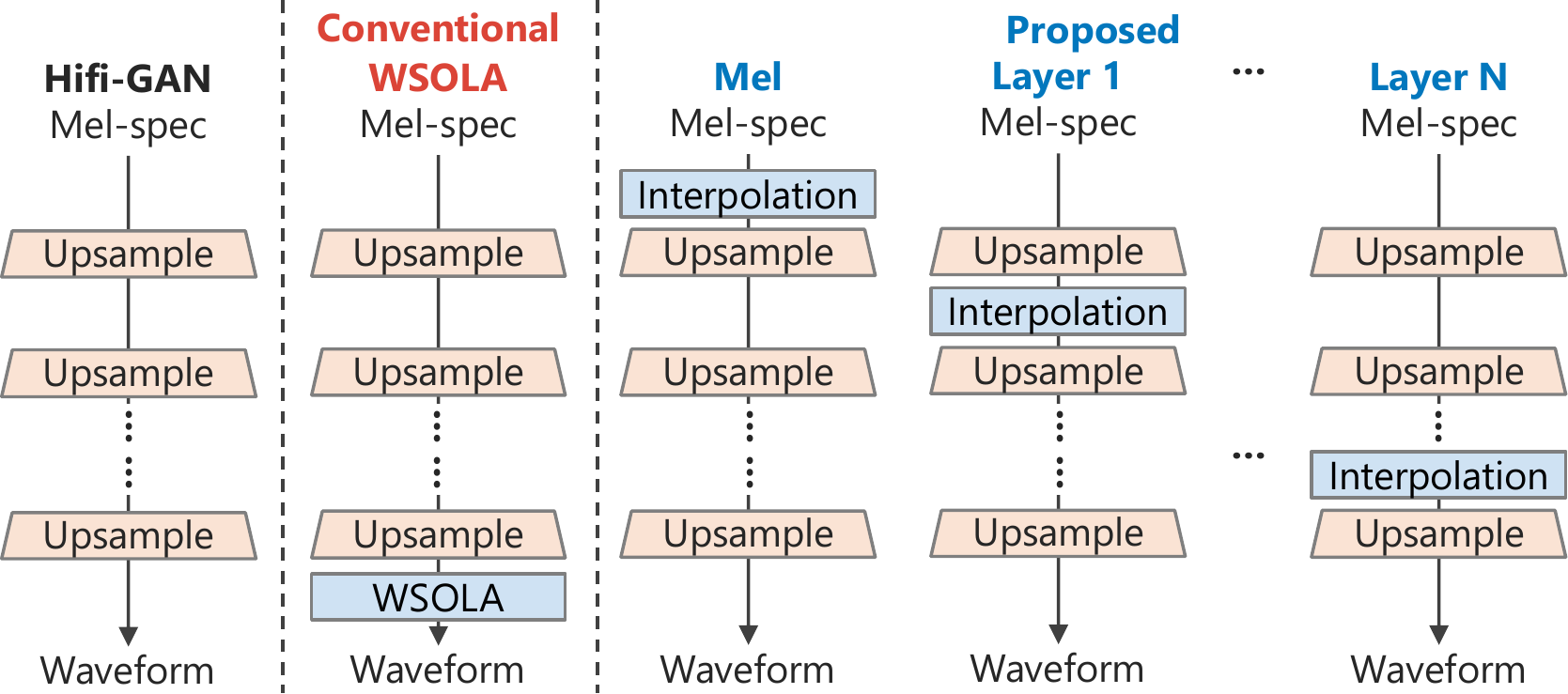}
  \caption{Diagram of our idea. An interpolation layer is inserted into HiFi-GAN to warp the length of mel-spectrograms or hidden features and control the speaking rate. In this work, we use a TSM algorithm WSOLA as the baseline method.}
  \label{fig:proposed}
  \vspace{-4mm}
\end{figure}

In this paper, we investigate the possibility of using a neural vocoder to control speaking rate.
We use a recent neural vocoder HiFi-GAN \cite{kong2020hifi} based on generative-adversarial-network (GAN) as vocoder, which can synthesize high-fidelity speech with $0.01$ of the real-time factor.
Our idea is illustrated in Figure~\ref{fig:proposed}.
The proposed method interpolates mel-spectrograms or hidden features in the possible inner layers of HiFi-GAN to control the speaking rate without fundamental frequency analysis as~\cite{okamoto2022spcom}.
When the interpolation layer is inserted before the generated waveform, the proposed method degrades to a TSM algorithm, which is used as the baseline method of this work.
We consider both a bandlimited signal resampling and an image scaling interpolation methods in the experiments.
To evaluate the proposed method, we built a corpus with three kinds of speaking rates of unique speakers and conducted experiments on it.
Results of comprehensive objective and subjective evaluations demonstrated that the proposed method can synthesize speech with higher fidelity and efficiency than the baseline TSM algorithm. 
The key contributions of this work are as follows:
\begin{enumerate} \leftskip -5mm \itemsep -0mm
    \item We propose a speaking rate control method by a neural vocoder with high fidelity and efficiency. The implementation is open-sourced in our website\footnote{\url{https://github.com/Aria-K-Alethia/speaking-rate-controllable-hifi-gan}}.
    \item We design a corpus for speaking rate control and give insights through experiments upon this corpus. The corpus is also open-sourced in our website\footnote{\url{https://ast-astrec.nict.go.jp/en/release/speedspeech_ja_2022/download.html}}. 
\end{enumerate}
% In this paper, we investigate the possibility of using a neural vocoder to control speaking rate. We use a recent GAN-based vocoder HiFi-GAN \cite{kong2020hifi} as the vocoder, which can synthesize high-fidelity speech with $0.01$ of the real time factor. Our idea is illustrated in \ref{fig:proposed}. The proposed method interpolates mel-spectrograms or hidden features in the possible inner layers of HiFi-GAN to control the speaking rate as~\cite{okamoto2022spcom}. We consider both a bandlimited signal resampling and an image scaling interpolation methods in the experiments.
% We also designs a corpus for speaking rate control. Our newly designed corpus consists of utterances that unique speakers utter unique texts with three kinds of speaking rates. Using this corpus, we can evaluate speed-controlled speech with actual speech that the speaker utters with the speaking rate. Results of evaluations using the designed corpus demonstrated that the proposed method can synthesize higher-fidelity speech while preserving computational efficiency. The key contributions of this work are as follows:
% \begin{enumerate} \leftskip -5mm \itemsep -0mm
%     \item We propose a speaking rate control method by a neural vocoder with high fidelity and efficiency. The implementation is open-sourced in our website\footnote{\url{https://github.com/Aria-K-Alethia/speaking-rate-controllable-hifi-gan}}. 
%     \item We design a corpus for speaking rate control and give insights through experiments upon this corpus. The corpus is also open-sourced\footnote{\url{https://ast-astrec.nict.go.jp/release/speedspeech_ja_2022/download.html}}. 
% \end{enumerate}
%2
\section{Proposed Method and Corpus}
\subsection{Method}
\subsubsection{Overview}
We use HiFi-GAN \cite{kong2020hifi} as the vocoder. Compared with the existing models~\cite{wn_vocoder,wavernn,waveglow,valin2019lpcnet,kumar2019melgan,pwg},
HiFi-GAN utilizes adversarial training to discriminate synthetic and genuine speech in parallel, thus can synthesize speech with high fidelity and efficiency.
As Figure~\ref{fig:proposed} illustrates, the basic idea of the proposed method is to use an interpolation layer to change data length so that the speaking rate of the synthesized speech will be different from the original mel-spectrogram\footnote{
    Yet another approach is to change upsampling intervals~\cite{morrison2021neural} and can be implemented by changing the stride width of the transposed convolution layers of HiFi-GAN. A preliminary experiment confirmed that the proposed method was better than this method.
}.
Since the generator of HiFi-GAN has several similar blocks to upsample the data, it is natural and intuitive to insert an interpolation layer between these blocks.

%As mentioned in the previous section, when the interpolation layer is inserted after the last upsampling layer, the proposed method becomes a traditional TSM algorithm.
We use the waveform similarity overlapping-add (WSOLA) algorithm \cite{verhelst1993overlap} as the baseline method.
WSOLA can change the speaking rate of the speech while maintaining the periodic patterns of the signal by finding a frame that has the maximal similarity to the current frame.
Benefiting from this property, WSOLA is suitable for modifying the time-scale of human speech.

\begin{figure}[t]
  \centering
  \includegraphics[width=0.65\linewidth]{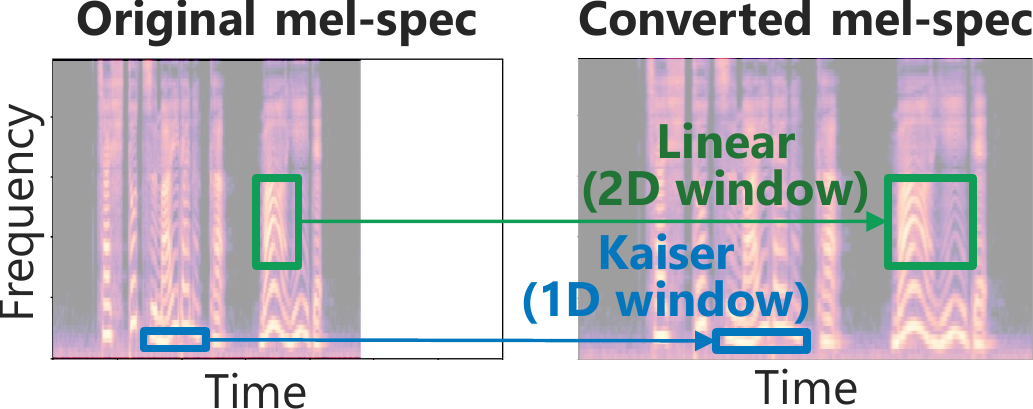}
  \caption{Examples of bandlimited resampling and linear interpolation. Bandlimited resampling is usually used to interpolate one-dimensional waveforms, while linear interpolation is designed to interpolate two-dimensional data like mel-spectrograms.}
  \label{fig:interpolation}
  \vspace{-5mm}
\end{figure}
\subsubsection{Interpolation Methods}
The interpolation layer should change the feature length while maintaining the semantic information within it.
In this work we consider two interpolation methods: bandlimited resampling based on the kaiser window~\cite{eldar2015sampling} and linear interpolation for image scaling.
Examples of these two methods are illustrated in Figure~\ref{fig:interpolation}.
Bandlimited interpolation is a commonly used resampling method in digital signal processing.
Since the outputs of the hidden layers of HiFi-GAN can be regarded as hidden representations of waveforms, we consider that it is reasonable to use this method to interpolate them.
Besides, it is also intuitive to interpret hidden features including mel-spectrograms as images that contain two axes time and frequency.
Therefore, we also use linear interpolation, a geometric interpolation method.
In the experiments we compare these two methods. %and show that linear interpolation is better than bandlimited resampling.

\subsection{Corpus}
To enhance the evaluation approaches of this task, we further design a speech corpus for speaking rate control.
The proposed corpus includes parallel texts spoken by a male and a female speaker with three kinds of speaking rates (slow, normal, fast).
The text is balanced in phonemes.
The speakers are professional so that all utterances of each speaking rate are intelligible.
%The speakers utter within a range of speaking rates that they can naturally speaker.
With this corpus, we can compare speaking-rate-controlled speech with ground-truth, e.g., comparing speech changed from ``slow'' to ``normal'' with actual ``normal'' speech. 
%3
\section{Experiments}
\subsection{Corpus Specification}
We hired a male and a female Japanese professional speaker to construct our corpus.
We used texts of the ITA corpus\footnote{\url{https://github.com/mmorise/ita-corpus}}, which is an open-sourced phoneme-balanced Japanese corpus.
Each speaker was required to speak $325$ utterances at three different rates: slow, normal, and fast, so each speaker has $975$ utterances.
We name this dataset \textit{SpeedSpeech-JA-2022}. Table~\ref{tab:speaking_rate} lists the exact speaking rate measured by mora per second.
This was computed by dividing the mora number by the voiced length of each utterance~\cite{mpm_2,mpm}.
The voiced frames were detected by using voice activity detection\footnote{\url{https://github.com/wiseman/py-webrtcvad}}.
It can be seen that the speaking rate of the fast speech is almost 2-time faster than the one of the slow speech.
Also, the male speaker speaks faster than the female speaker.

\begin{table}[t]
    \centering
    \caption{Speaking rate (mora per second) of the proposed SpeedSpeech-JA-2022 dataset.}
    \label{tab:speaking_rate}
    %\begin{tabular}{p{0.5cm}<{\centering}p{0.7cm}<{\centering}p{0.6cm}<{\centering}p{1.2cm}<{\centering}p{0.6cm}<{\centering}p{1.2cm}<{\centering}p{0.5cm}<{\centering}}
    \begin{tabular}{cccc}
    \hline
    Speaker & Fast & Normal & Slow \\
    \hline \hline
    Male    & $9.0 \pm 0.3$ & $6.4 \pm 0.4$ & $5.0 \pm 0.2$ \\
    Female  & $6.8 \pm 0.6$ & $4.8 \pm 0.3$ & $3.8 \pm 0.3$ \\
    Average & $7.9 \pm 1.2$ & $5.6 \pm 0.9$ & $4.4 \pm 0.7$ \\
    \hline
    \end{tabular}
\end{table}

\subsection{Experimental Setup}
We randomly picked up $45$ utterances for each speaker in which each rate has $15$ utterances.
The texts of the $15$ utterances of each speaker between different rates were kept to be the same so that we could measure the mel-cepstral distortion in the later experiments.
For all speech of each speaking rate, we converted them to the rest two rates and regarded the speech of the target rate as the ground-truth data.
Since every utterance has a different speaking rate, we computed the conversion factor for each utterance separately.
Denoting the durations of the source and the target utterances with the same text as $t_{\mathrm{src}}$ and $t_{\mathrm{tgt}}$, the conversion factor $f$ is defined as ${t_{\mathrm{src}}}/{t_{\mathrm{tgt}}}$.
In the subjective evaluations, we also converted the speech with standard factors $\{0.25, 0.50, 0.75, 1.25, 1.50, 1.75, 2.00\}$ that are widely used in real-world applications.

We used a pretrained universal HiFi-GAN model\footnote{\url{https://github.com/jik876/hifi-gan}} to convert mel-spectrograms into time-domain waveforms.
This model was trained on multilingual datasets so it can synthesize Japanese without further training.
Since the generator of the pretrained HiFi-GAN has four upsampling layers, it is possible to insert the interpolation layer into $5$ places (mel-spectrogram and layer $1, 2, 3, 4$).
All combinations of interpolation methods and places were investigated.
As a result, there were $11$ methods including $10$ proposed methods ($5$ places and $2$ interpolation methods) and $1$ baseline method in the experiments.
% For simplicity, we denote these methods as mel (l/k), L\{1,2,3,4\} (l/k) and WSOLA, where l denote linear interpolation and k denote bandlimited resampling based on kaiser window.
We used the bandlimited resampling class in torchaudio\footnote{\url{https://pytorch.org/audio/stable/transforms.html#torchaudio.transforms.Resample}} and the linear interpolation function of pytorch\footnote{\url{https://pytorch.org/docs/stable/generated/torch.nn.functional.interpolate.html}} as the implementations.
For the WSOLA algorithm we used audiotsm\footnote{\url{https://github.com/Muges/audiotsm}} package as the implementation.

% TTS 評価
In addition, we also evaluated the proposed method with a speech synthesis model.
We trained text-to-speech (TTS) models FastSpeech2~\cite{ren2020fastspeech} followed by the aforementioned universal HiFi-GAN model.
We followed the model architecture and hyperparameters of the open-sourced implementation\footnote{\url{https://github.com/ming024/FastSpeech2}}.
We trained a FastSpeech2 model using the JSUT corpus~\cite{jsut-jvs} and fine-tuned it using the SpeedSpeech-JA-2022 corpus.
The fine-tuning was performed with $10$k steps. The pretrained model was fine-tuned on the utterances of different speakers and speaking rates separately, i.e., there were six FastSpeech2 models in total.
The test data size was the same as above, and the remaining data were used for fine-tuning.

\subsection{Objective Evaluations}
\subsubsection{Mel-cepstral Distortions}
We first measured mel-cepstral distortions (MCDs) between the converted and ground truth speech to evaluate each method.
Utterance pairs were aligned by dynamic time warping.
The result is shown in Table~\ref{tab:mcd}.
The mel-spectrogram linear interpolation method (``Linear" and ``mel") obtained the best performance.
Besides, it can be observed that interpolating mel-spectrograms is better than interpolating hidden features of inner layers.
The baseline method (WSOLA) also obtained relatively good performance but is still worse than the two mel-spectrogram interpolation methods.
Surprisingly, all the methods of interpolating hidden features of inner layers have poor performance.
Here we cannot list all results due to the page limit, but the MCDs degraded as the interpolation layer was placed close to the output layer.
After a preliminary analysis, we found that the fundamental frequency (F0) of the utterances converted by these methods were destructed, though the semantic information was well preserved.
We assume this is because the convolutional neural network architecture of HiFi-GAN makes it depend on the hidden feature length to work, thus changing the feature size will influence the behavior without further training.

% \begin{table*}[t]
% \centering
% \caption{Mel-cepstral distortion of the converted speech. HiFi-GAN represents synthesized speech without rate changing.}
% \label{tab:mcd}
% \begin{tabular}{cccccccccccccc}
% \hline
% From & To   & HiFi-GAN & mel (l) & mel (k) & WSOLA & L1 (l) & L2 (l) & L3 (l) & L4 (l) & L1 (k) & L2 (k) & L3 (k) & L4 (k) \\
% \hline \hline
% Slow & Norm & 1.15 & \textbf{2.20} & 2.22 & 2.37 & 3.38 & 2.74 & 3.12 & 3.39 & 3.35 & 2.72 & 3.14 & 3.40 \\
% Slow & Fast & 1.15 & 2.99 & \textbf{2.96} & 3.18 & 3.85 & 4.07 & 4.38 & 4.92 & 3.81 & 4.06 & 4.40 & 4.93 \\
% Norm & Slow & 1.16 & \textbf{2.24} & \textbf{2.24} & 2.37 & 3.42 & 2.85 & 3.05 & 3.27 & 3.46 & 2.85 & 3.06 & 3.31 \\
% Norm & Fast & 1.16 & \textbf{2.74} & 2.76 & 2.91 & 3.80 & 3.54 & 3.88 & 4.13 & 3.83 & 3.53 & 3.89 & 4.15 \\
% Fast & Slow & 1.41 & \textbf{2.83} & 2.87 & 3.03 & 3.49 & 3.68 & 3.75 & 3.94 & 3.48 & 3.67 & 3.76 & 3.98 \\
% Fast & Norm & 1.41 & \textbf{2.62} & 2.65 & 2.83 & 3.62 & 3.32 & 3.55 & 3.78 & 3.65 & 3.33 & 3.56 & 3.82 \\
% \hline
% \end{tabular}
% \end{table*}

\begin{table}[t]
\setlength{\tabcolsep}{1mm}
\centering
\caption{Mel-cepstral distortion of the converted speech. ``HiFi-GAN'' represents synthesized speech without rate changing. ``\#'' (1--4) indicates the layer index of the best performance. \textbf{Bold} indicates the best score.}
\label{tab:mcd}
\footnotesize
\begin{tabular}{cc||c|c|cc|cc}
\multicolumn{2}{c||}{Speed} &           &       & \multicolumn{2}{c|}{Linear} & \multicolumn{2}{c}{Kaiser} \\ 
From        & To          & Hifi-GAN    & WSOLA & mel           & layer (\#)    & mel           & layer (\#)     \\ \hline
Slow        & Norm        & 1.15        & 2.37  & \textbf{2.20} & 2.74 (2)      & 2.22          & 2.72 (2)                 \\
Slow        & Fast        & 1.15        & 3.18  & 2.99          & 3.85 (1)      & \textbf{2.96} & 3.81 (1)                 \\
Norm        & Slow        & 1.16        & 2.37  & \textbf{2.24} & 2.85 (2)      & \textbf{2.24} & 2.85 (2)                 \\
Norm        & Fast        & 1.16        & 2.91  & \textbf{2.74} & 3.54 (2)      & 2.76          & 3.53 (2)                 \\
Fast        & Slow        & 1.41        & 3.03  & \textbf{2.83} & 3.49 (1)      & 2.87          & 3.48 (1)                 \\
Fast        & Norm        & 1.41        & 2.83  & \textbf{2.62} & 3.32 (2)      & 2.65          & 3.33 (2)                
\end{tabular}
\end{table}

\subsubsection{Real-Time Factors}
\begin{figure}[t]
  \centering
  \includegraphics[width=0.98\linewidth]{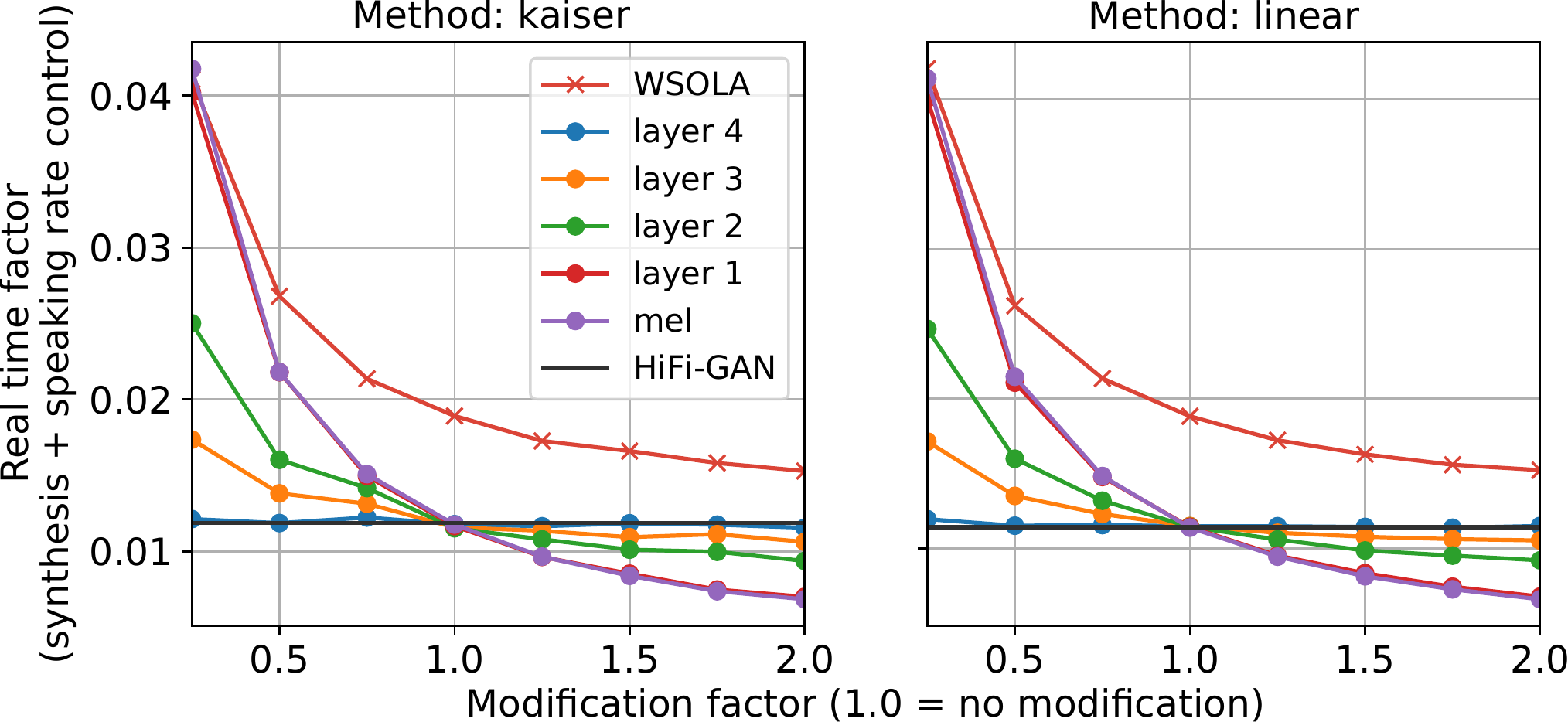}
  \caption{Results of efficiency evaluation. ``HiFi-GAN'' represents the efficiency of the model itself without changing the speaking rate.}
  \label{fig:rtf}
  \vspace{-3mm}
\end{figure}

To evaluate the efficiency of each method, we then computed real-time factors (RTFs) for each method using an NVIDIA GeForce RTX 2080 Ti GPU card.\footnote{HiFi-GAN can also realize real-time inference on CPUs~\cite{kong2020hifi,asru2021okamoto}.}
The computation time is defined as the summation of the generation time of HiFi-GAN and the speaking rate conversion time of each method.
For simplicity in this evaluation we only used standard factors to convert all slow, normal, and fast utterances but did not convert between them.
The result is shown in Figure~\ref{fig:rtf}.
It can be seen that all the proposed methods have better efficiency than the baseline WSOLA method.
The mel-spectrogram interpolation methods which obtained the best performance in the MCDs evaluation have the best performance when the conversion factor is greater than $1.0$ but obtained the worst performance among the proposed methods when the conversion factor is less than $1.0$.
This is because the four upsampling blocks in HiFi-GAN will magnify the data length changing effect, therefore the shortened or elongated data will become shorter or longer after being processed by the blocks, and further increase or reduce the computation time.

% \begin{figure}[t]
%   \centering
%   \subfigure[Bandlimited resampling]{
%         \centering
%         \includegraphics[width=0.5\linewidth]{figures/rtf_kaiser.pdf}
%   }% sub-figure1
%   \subfigure[Linear interpolation]{
%         \centering
%         \includegraphics[width=0.5\linewidth]{figures/rtf_linear.pdf}
%   }% sub-figure2
%   \caption{Results of efficiency evaluation. The efficiency measured by RTF is compared for (a) proposed method using bandlimited resampling and WSOLA, (b) proposed method using linear interpolation and WSOLA. HiFi-GAN represents the efficiency of the model itself without changing the speaking rate.}
%   \label{fig:rtf}
% \end{figure}

\subsection{Subjective Evaluations}
% \subsubsection{Speech Naturalness of HiFi-GAN}
% \label{subsubsection:hifi_naturalness}
% We first conducted a standard 5-scale mean opinion score (MOS) test to evaluate the synthetic speech generated by HiFi-GAN without rate control.
% Totally 100 listeners joined in the test.
% Each utterance was evaluated by 10 listeners.
% The result is shown in~\ref{fig:hifi-gan_mos}.
% \begin{figure}[t]
%   \centering
%   \includegraphics[width=0.95\linewidth]{figures/hifi-gan.pdf}
%   \caption{Results of naturalness MOS evaluation for natural speech (GT) and synthetic speech (HiFi-GAN) of each speaker under different speaking rate without speaking rate control.}
%   \label{fig:hifi-gan_mos}
% \end{figure}
% \begin{figure*}[t]
%   \centering
%   \includegraphics[width=0.95\linewidth]{figures/cer.pdf}
%   \caption{Results of intelligibility evaluation. The title of each sub-figure represents the speaker and the SNR. The red vertical line represents speech without changing the speaking rate.}
%   \label{fig:cer}
% \end{figure*}
% It can be seen that the listeners tend to rate low scores for speech with uncommon speaking rates, even for the natural speech.
% Also, there is still a quality gap between the natural and synthetic speech, which we assume is because we didn't fine-tune the model on the dataset.

\subsubsection{Speech naturalness: fixed source rates}
In the subjective evaluations we evaluated the naturalness of the converted speech by mean opinion score (MOS) tests~\cite{p830}.
On the basis of the results of the objective evaluations, we only selected three methods: WSOLA, mel-spec interpolation using linear interpolation and bandlimited resampling with kaiser window in this test.

\begin{figure}[t]
  \centering
  \includegraphics[width=0.98\linewidth]{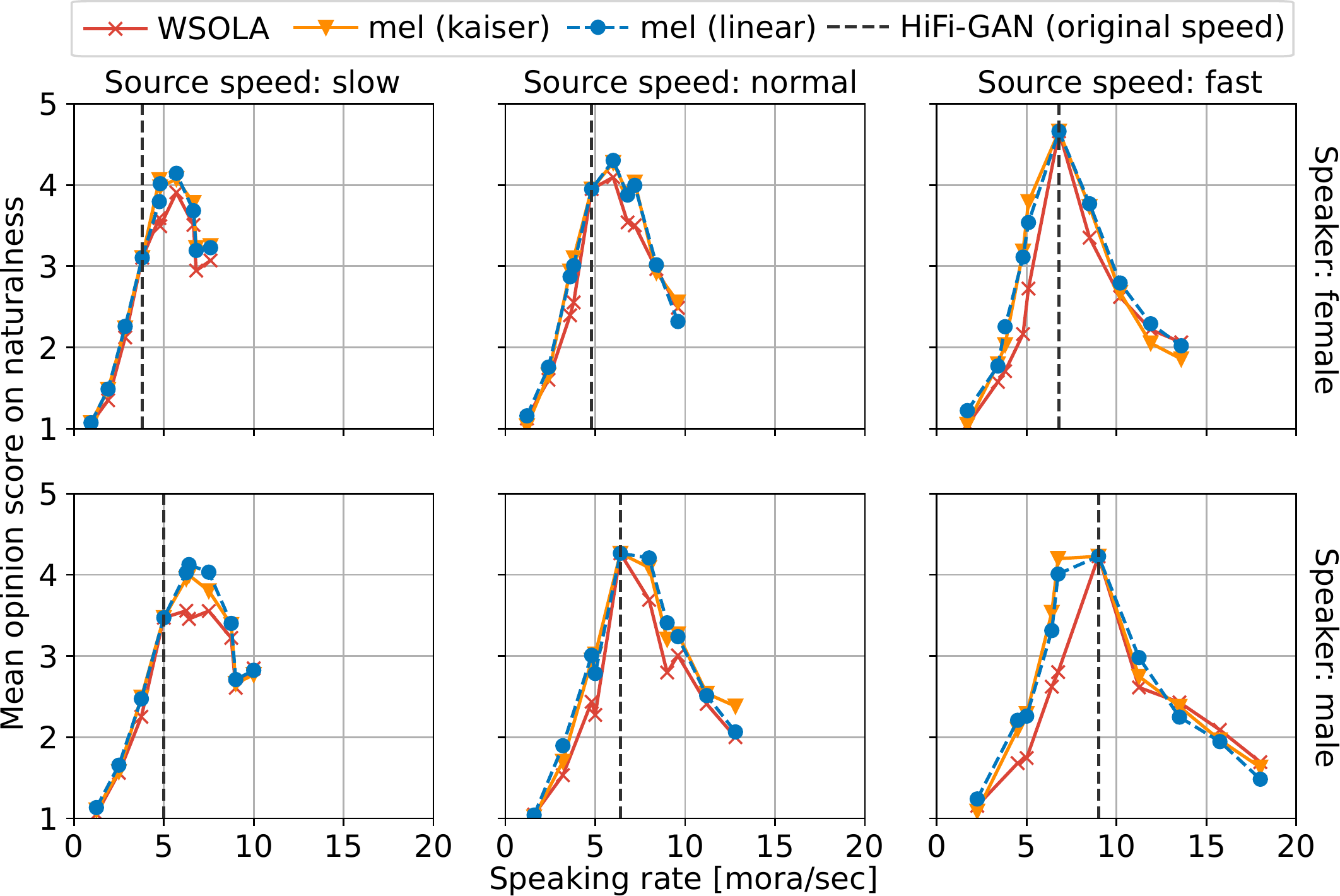}
  \caption{Results of naturalness MOS evaluation.}
  \label{fig:mos}
  \vspace{-5mm}
\end{figure}

In the first MOS test we intended to clarify the influences of the source speaking rates on the MOS.
Totally $1,500$ listeners joined in the test.
Each utterance was evaluated by $30$ listeners.
The result is illustrated in Figure~\ref{fig:mos}.
It can be observed that if the speaking rate is in a common range $[4, 10]$, the two mel-spectrogram interpolation methods are better than WSOLA and the linear interpolation method is slightly better than the kaiser window method, which is consistent with the conclusions we got from the objective evaluations.
This implies that it is easy to slow down the utterances with high speaking rates or speed up the utterances with low speaking rates.
Second, when the speaking rates become extremely slow or fast, the MOS scores become very low and it is hard to distinguish which method is better for the listeners.
We assume that there are two possible reasons for this problem.
One possible reason is that the pretrained HiFi-GAN model was trained on a corpus with normal speaking rates, so it cannot synthesize speech well with slow or fast speaking rates.
Another possible reason is that the listeners are not used to listening to speech with uncommon speaking rates.
But since in the later experiments it can be observed that the listeners rated low scores even for natural speech with uncommon speaking rates, (Section~\ref{subsubsection:comp_with_gt}), we conclude that the latter reason is more appropriate to explain the result.

\subsubsection{Speech naturalness: fixed target rates}
\label{subsubsection:comp_with_gt}
% In the second evaluation, we instead aimed to know the performance of the proposed method when the target speaking rate is fixed.
% Based on the first evaluation, we only used the utterances converted by the mel (l) method.
% For simplicity we did not use the utterances converted using standard factors.
% We categorized all utterances into 6 groups by their speaker and target speaking rate, and conducted a MOS test for each group.
% Each test has 35 listeners.
% The result is illustrated in~\ref{fig:mos_per_speed}.
% It can be seen that when the source and target speaking rates are quite different, e.g. slow to fast, the performance becomes bad.
% Also, the listeners tend to rate high scores for utterances with normal speaking rates, which is consistent with the conclusion of the previous evaluation.
% In the task of converting female slow utterances to normal utterances, the performance of the proposed method even surpasses HiFi-GAN, which demonstrates the effectiveness of the proposed method.
% Although when the target speaking rate is slow or fast, the performance of the proposed method is not as good as HiFi-GAN, in the later section we show that the proposed method can obtain a low character error rate in an intelligibility test, which demonstrates that the listeners can at least understand the converted utterances even if they are not used to listening to such speech.
In the second test we instead aimed to know the performance of the proposed method compared to the ground truth under specified speaking rates.
Based on the first evaluation, we only used the proposed method using mel-spectrogram linear interpolation.
We specified three speaking rates (slow, normal, fast) for each speaker.
For each specification, we prepared four types of speech: ground-truth raw speech (i.e., the speaker's natural speech with the specified speaking rate); the analysis-synthesized speech using HiFi-GAN without speaking rate control; and two kinds of rate-controlled speech converted from the rest two speaking rates, e.g. ``From normal" and ``From fast" for slow speech.
%The last was made from data of other speaking rates, e.g., ``slow'' and ``fast'' speech were converted to ``normal'' when the specified rate was ``normal.''
We conducted 6 MOS tests on naturalness for all combinations of speakers and speaking rates. 
Each test had $35$ listeners; each listener evaluated $20$ samples.

\begin{figure}[t]
  \centering
  \includegraphics[width=0.98\linewidth]{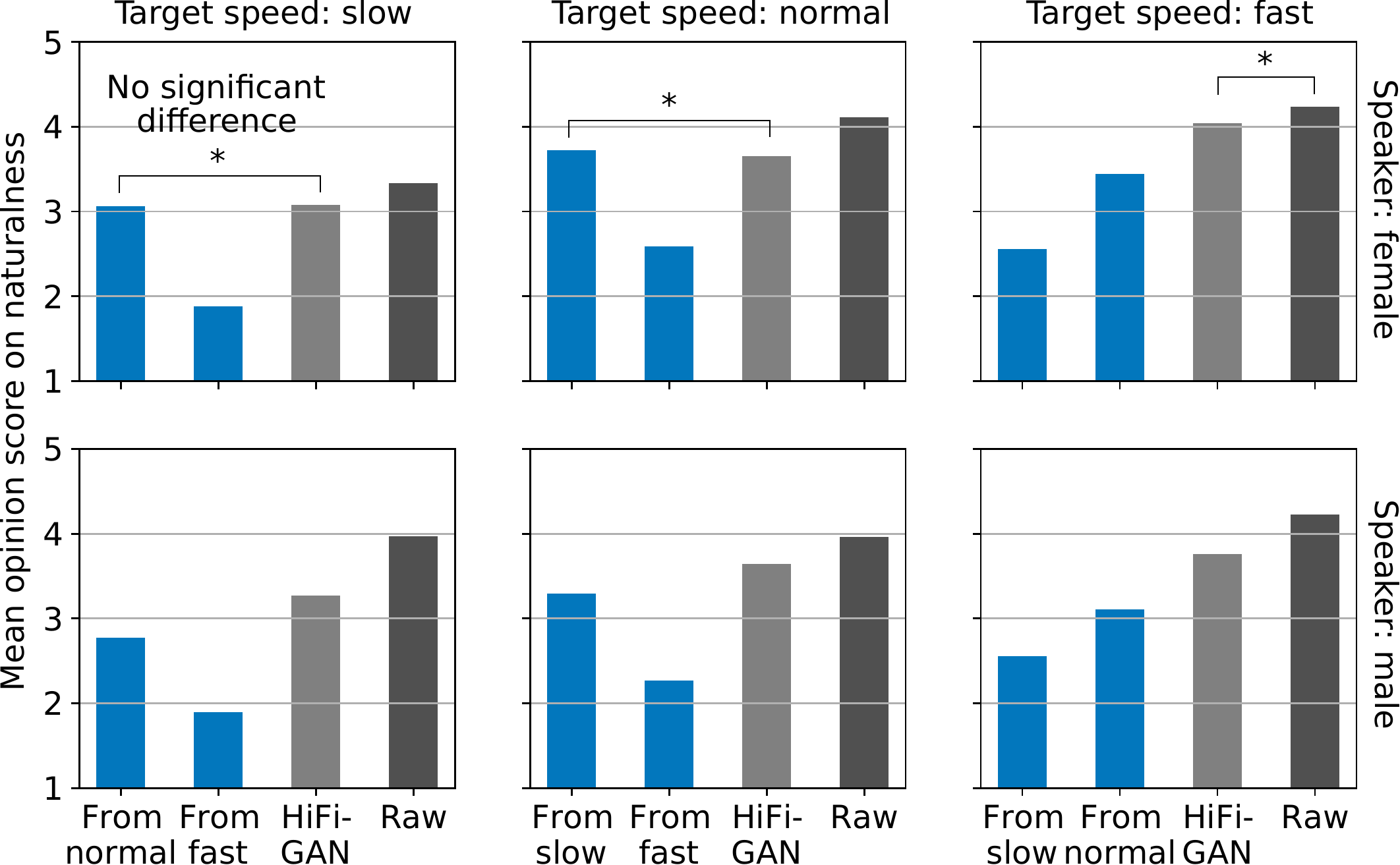}
  \caption{Results of naturalness MOS evaluation using analysis-synthesized speech with fixed target speaking rate. ``HiFi-GAN'' represents speech without rate changing. Scores with ``*'' are not significantly different, i.e., $p > 0.05$.}
  % \caption{Results of naturalness MOS evaluation with fixed target speaking rate. The title of each sub-figure represents the speaker and the target speaking rate. The red bar denotes the mel (l) method which obtained the best performance in the previous evaluations.}
  \label{fig:mos_per_speed}
\end{figure}

Figure ~\ref{fig:mos_per_speed} shows the result.
It can be seen that when the source and target speaking rates are quite different, e.g. slow to fast, the performance becomes bad.
Also, the listeners tend to rate high scores for utterances with normal speaking rates, which is consistent with the conclusion of the previous evaluation.
In the task of normal-to-slow and slow-to-normal conversion of the female speaker's speech, the performance of the proposed method even surpasses HiFi-GAN, which demonstrates the effectiveness of the proposed method.
However, when the target speaking rate is slow or fast, the performance of the proposed method is not as good as HiFi-GAN.
%This is the remaining issue of the speaking rate control.

Finally, we conducted a test using TTS-synthesized speech.
The speaking rate control function was added to the speech synthesis system.
Specifically, mel-spectrograms output from FastSpeech2 were converted into waveforms by HiFi-GAN after being interpolated.
The experimental setup is almost the same as above except that (1) we use TTS-synthesized speech without speaking rate control to replace ``HiFi-GAN" and (2) add WSOLA for comparison.
Each test had $35$ listeners; each listener evaluated $25$ samples.

\begin{figure}[t]
  \centering
  \includegraphics[width=0.98\linewidth]{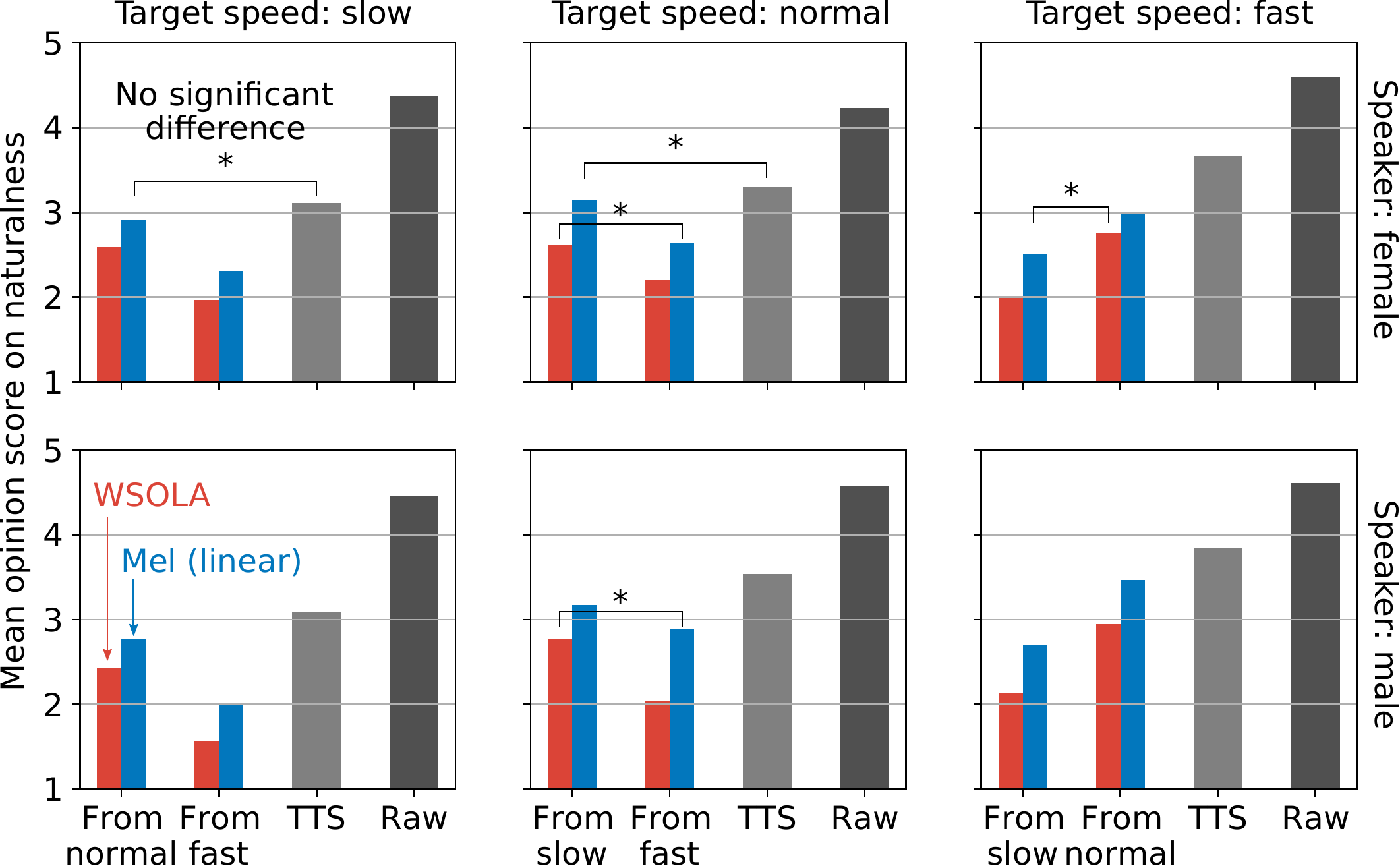}
  \caption{Results of naturalness MOS evaluation using TTS speech with fixed target speaking rate. ``TTS'' represents speech without rate changing. Scores with ``*'' are not significantly different, i.e., $p > 0.05$.}
  % \caption{Results of naturalness MOS evaluation with fixed target speaking rate. The title of each sub-figure represents the speaker and the target speaking rate. The red bar denotes the mel (l) method which obtained the best performance in the previous evaluations.}
  \label{fig:mos_per_speed_tts}
  \vspace{-3mm}
\end{figure}

Figure~\ref{fig:mos_per_speed_tts} shows the result.
First, the proposed method always outperforms the baseline.
The overall tendencies are similar to the result of the previous test.
The normal-to-slow and slow-to-normal conversion of the female speaker's speech are comparable to slow TTS and normal TTS, respectively.
In other cases, the rate-controlled speech is still worse than the speech without rate control.
%This is also the remaining issue of the speaking rate control.

\section{Conclusions}
This paper described a method for speaking rate control by HiFi-GAN using feature interpolation.
The idea of the proposed method is to insert a feature interpolation layer into the model to change the data length and control the speaking rate.
A Japanese corpus with three speaking rates was designed to evaluate the proposed method.
Results of subjective and objective evaluations demonstrated that the proposed method of mel-spectrogram interpolation using linear interpolation had better efficiency and quality performance than the baseline TSM algorithm WSOLA.
The future work will be testing the generalization ability of the proposed method on other neural vocoders.

\small{
\textbf{Acknowledgements:}
This work is supported by JST SPRING, Grant Number JPMJSP2108, and is also supported by JSPS KAKENHI 19H01116, 21H04900.
}

\bibliographystyle{IEEEtran}

\bibliography{mybib}

\end{document}